\documentclass[aps,prb,twocolumn,superscriptaddress,floatfix]{revtex4-2}
\usepackage[utf8]{inputenc}
\usepackage[english]{babel}
\usepackage{amsmath}
\usepackage{feynmf}
\usepackage{xcolor}
\usepackage{soul}
\usepackage{graphicx}
\usepackage{amsfonts}
\usepackage{blindtext}
\usepackage{appendix}
\usepackage[colorlinks=true,unicode=true,allcolors=blue]{hyperref}
\usepackage{siunitx}

\usepackage[normalem]{ulem}

\renewcommand{\vec}[1]{\mathbf{#1}}
\newcommand{\vk}{{\vec k}}
\newcommand{\vq}{{\vec q}}
\newcommand{\eps}{{\varepsilon_\vk}}
\newcommand{\E}{{E_\vk}}
\newcommand{\D}{{\Delta_\vk}}

\begin{abstract}
Cuprates are \textit{d}-wave superconductors which exhibit a rich phase diagram: they are characterized by superconducting fluctuations even above the critical temperature, and thermal disorder can reduce or suppress the phase coherence. However, photoexcitation can have the opposite effect: recent experiments have shown an increasing phase coherence in optimally doped BSCCO with mid-infrared driving. Time-resolved terahertz spectroscopies are powerful techniques to excite and probe non-equilibrium states of superconductors, directly addressing collective modes, such as amplitude (Higgs) oscillations. In this work, we calculate the full time evolution of the current generated by a cuprate with a quench-drive spectroscopy setup. Analyzing the response in Fourier space with respect to both the real time and the quench-drive delay time, we look for the signature of a transient modulation of higher harmonics as well as the Higgs mode, in order to characterize the ground state phase. 
In particular, this approach can provide a smoking gun for induced or increased phase coherence when applied to the pseudogap phase. These results can pave the way for future experimental schemes to characterize and study superconductors alongside incoherent phases and phase transitions, including induced and transient superconductivity.
\end{abstract}

\begin{document}

\title{Quench-drive spectroscopy of cuprates}
\author{Matteo Puviani}
\email{m.puviani@fkf.mpg.de}
\affiliation{Max Planck Institute for Solid State Research, 70569 Stuttgart, Germany}
\author{Dirk Manske}
\email{d.manske@fkf.mpg.de}
\affiliation{Max Planck Institute for Solid State Research, 70569 Stuttgart, Germany}

\date{\today}
\maketitle

\section{Introduction} 
High-temperature superconductors have attracted much of the attention on the research on superconductivity since their discovery.
In the last years, experiments on high-temperature superconductors have revealed the presence of signatures of superconducting fluctuations above the critical temperature $T_c$. This behaviour has been attributed to incoherent or pre-formed Cooper pairs, leading to third-harmonic generation (THG) and enhancement of the reflectivity change in a pump-probe experimental configuration \cite{PhysRevLett.122.067002,chu2020phase,chu2021fano}.

Cuprates are the prototypical example of high-temperature superconductors, exhibiting a complex phase diagram as a function of doping and temperature, with a superconducting dome enriched by charge-density wave and pseudogap phases \cite{Keimer2015}.
In fact, optimally doped Y-Bi2212 exhibits a superconducting phase below $T_c = 97$ K, and a pseudogap phase above $T_c$ and up to the temperature $T^* \approx 135$ K \cite{PhysRevB.88.245132,PhysRevB.95.024501,PhysRevLett.122.067002,PhysRevB.104.125121,Abdullah_Kaplan2021-zv}. It has been argued that the pseudogap phase on top of the superconducting dome at optimal doping in unconventional superconductors is driven by the loss of phase coherence between the Cooper pairs, rather than the softening or vanishing of the pairing strength \cite{PhysRevB.81.054510,PhysRevB.104.214510, singh2021fermi}. Therefore, in the pseudogap phase the electrons are still paired, but their local phase is different from the global phase of the order parameter, which is lowered as a consequence. 

A variety of non-equilibrium experiments on cuprates have indicated the importance and the interaction between between collective modes, such as the amplitude (Higgs) mode and Josephson plasmon: different setups have been used to investigate it, from high-harmonic generation to pump-probe spectroscopy, to reflectivity measurements  \cite{TsujiHiggs,PhysRevX.11.011055,Gabriele2021,chu2020phase,chu2021fano}.
Moreover, it has been recently shown that THz pulses in the mid-infrared region can dynamically enhance the phase coherence of Cooper pairs in optimally doped cuprates, which is lowered by thermal disorder in equilibrium conditions \cite{PhysRevB.104.125121}.

\begin{figure}[h!]
\centering
\includegraphics[width=8cm]{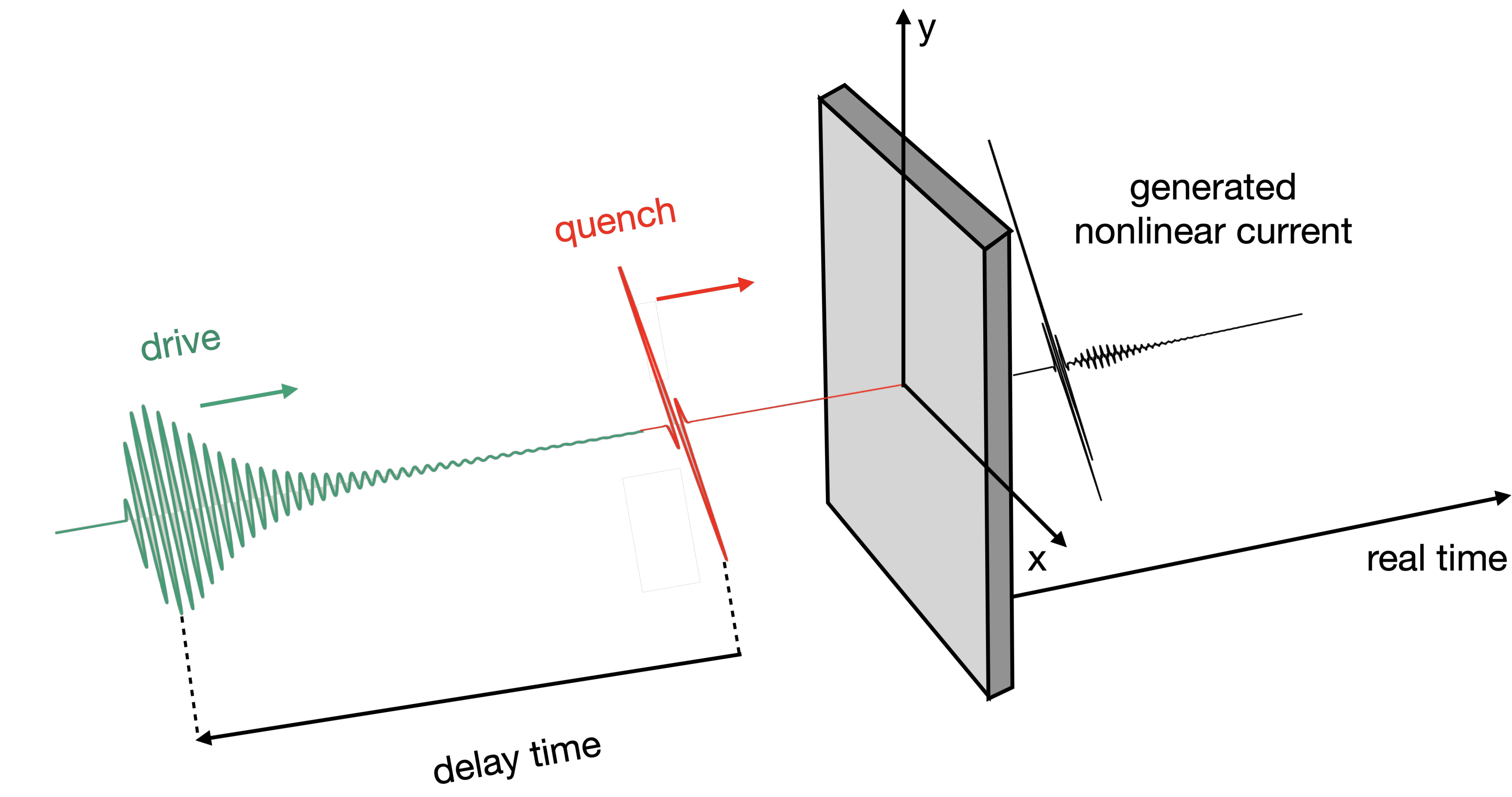}
\caption{\textbf{Quench-drive setup.} The figure shows the scheme of the quench-drive setup used here: the quench pulse (red) impinges on the material (grey). Then, after a a delay time ($\Delta t$, measured from the maximum peak of the quench to the maximum peak of the drive) the driving pulse also interacts with the material: the nonlinear current output is then generated and can be detected in real time.} \label{scheme1}
\end{figure}

In this paper we go beyond previous works of pump-probe spectroscopy on cuprates, applying the quench-drive spectroscopy setup \cite{Woerner2013,PhysRevLett.118.207204,PhysRevLett.122.257401,Mahmood2021}  (see Fig.\ref{scheme1}) recently extended to study different features of superconductors \cite{puviani2021transient}, to superconductors with anisotropic d-wave order parameter. In this configuration, a first few-cycle THz broadband pulse (quench, red in Fig.~\ref{scheme1}) impinges on the material, followed after an adjustable delay time $\Delta t$ by a multi-cycle THz narrow-band (asymmetric) driving pulse (drive, green in Fig.~\ref{scheme1}). Then, the generated current, which is experimentally addressed by measuring the change of the transmitted electric field or the nonlinear optical conductivity \cite{PhysRevLett.108.097401,Demsar2020}, can be analyzed in transmission in real time after filtering out the linear response directly proportional to the incoming pulses. 
Quench-drive spectroscopy has been shown to be a versatile and powerful tool to systematically analyze the superconducting response as well as characterizing the signatures of collective modes, possibly enhancing the overall measured signal \cite{puviani2021transient}. 
This method goes beyond standard pump-probe spectroscopy, where a short-time pulse drives the system out of equilibrium, followed by a subsequent weak and short-time pulse which probes the system (e.g.: current, reflectivity) \cite{PhysRevLett.120.117001,LI201529}. In a quench-drive setup, in fact, both the short-time pulse and the long-time drive can perturb the system, and the non-fixed relative geometry of the two pulses allows to shift the quench-drive time delay. This provides an extended number of possible configurations, allowing for the quench to overlap with the drive, or even acting on the driven superconductor after the longer pulse. The analysis of the generated nonlinear current in both time and Fourier space, for example, including not only the real evolution but also the time-dependence of the quench and drive relative time, allows to obtain a richer signal than the usual harmonic response of the non-equilibrium material. In fact, the application of this spectroscopic technique to conventional clean $s$-wave superconductors has demonstrated to provide the nonlinear current signal due to the quasiparticles' and collective modes' excitations, visible as distinctive features in the two-dimensional time and frequency plots, respectively \cite{puviani2021transient}. Alongside the high-harmonic generation, both transient excitation and dynamical modulation of the generated harmonics are visible, and can be theoretically described by solving the real-time Heisenberg's equation of motion, or interpreted with a diagrammatic approach. \\
However, in quench-drive spectroscopy the real and delay times are not fully independent, since they both refer to the driving pulse, and are not able to catch the decoherence processes. This mainly differs with other three-pulses techniques, like pump-dump-probe, pump-push-probe or pump-repump-probe mechanisms, which have been widely used to study molecular excitations and transient states since decades \cite{Gai1997,vanWilderen2009,Fitzpatrick2012,Kee2014}. In pump-repump-probe spectroscopy, for example, the first pulse creates a macroscopic polarization which decays due to dephasing, the second one induces a population of the excited state, while the probe converts it back to a coherent polarized state \cite{Giannetti2016}.
An extension of our scheme to a real two-dimensional coherent spectroscopy, with two independent times and probing of decoherence processes, will be object of a future work.

\begin{figure*}[htb]
\centering
\includegraphics[width=16cm]{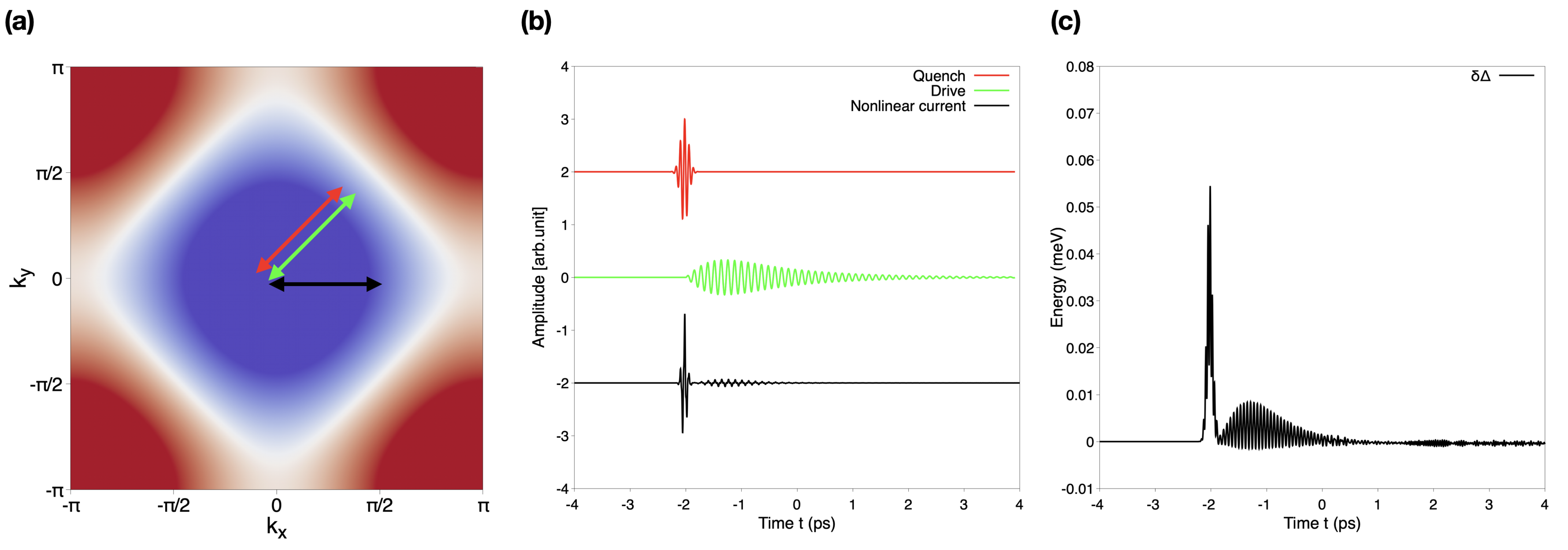}
\caption{\textbf{Quench-drive spectroscopy.} \textbf{(a)} Band structure in the first Brillouin zone used to reproduce the optimally doped Bi2212: the red and green arrows represent the direction of linear polarization of the quench and drive pulses, respectively. The black arrow indicates the direction of the measured current, i.e. along the $x$ axis. \textbf{(b)} The figure shows the external vector potential of the quench (red) and the drive (green) for an exemplary case with fixed quench-drive delay time, as well as the nonlinear current response (black). \textbf{(c)} Time-dependent superconducting order parameter variation, $\delta \Delta (t) = \Delta (t) - \Delta (0)$, due to the quench and drive pulses in (b).} \label{plot}
\end{figure*}

In the present work, we numerically solve the Heisenberg's equation of motion derived within the pseudospin formalism \cite{Science.345.1145,PhysRevB.92.064508} in order to describe the superconducting state, and we support our interpretations and results with the derivation of the nonlinear susceptibility by means of a diagrammatic approach \cite{PhysRevB.93.180507}. Moreover, in order to treat the pseudogap phase, characterized by incoherent pairs, we extend the pseudospin model artificially adding a phase to the Cooper pairs \cite{PhysRevLett.122.067002}, and solving the corresponding equations to obtain the time-evolution of the order parameter and the generated nonlinear current (see Fig.\ref{plot}(b),(c)). 

The paper is organized as follows: in Section \ref{theory} we describe the theoretical models that we used. Starting from the pseudospin approach to solve the equations of motion in order to calculate the time-dependent superconducting gap and the generated nonlinear current, we extend it in order to be able to describe incoherent pairs. In addition, we perform the same calculations for the superconducting state using a diagrammatic approach, deriving the nonlinear susceptibility responsible for the measured response.
In Section \ref{results} we show the results of the numerical experiments for the quench-drive setup on a cuprate: we analyze the two dimensional plots in time and frequency domain, detecting the presence of higher harmonics and signatures of quasiparticles' and the Higgs mode. Then, we repeat the calculation in the presence of incoherent pairs: we show that even a moderate incoherence which only slightly reduces the superconducting gap can suppress the high-harmonic generation in the nonlinear current, as well as the quasiparticles' and amplitude mode's response. 
Finally, we provide a summary and an outlook on future applications and perspectives of quench-drive spectroscopy in Section \ref{conclusion}.

\section{Theoretical background} \label{theory}

In this section we formulate the theoretical approach used to investigate the current generated by a clean (high-temperature) superconductor subject to an external field. We first develop the pseudospin model for a general superconductor, then we extend it in order to be able to describe incoherent pairs. Finally, we show a diagrammatic approach, where we derive the nonlinear susceptibility used to obtain the nonlinear current, disentangling quasiparticles' and Higgs' contributions. 

\subsection{Pseudospin model approach} \label{sec:pseudo}
In order to describe the superconducting phase of a material, we adopt the BCS model expressed by the mean field Hamiltonian
\begin{align} \label{BCSHam}
H_{BCS} = \sum_{\vk, \sigma} \eps \hat{c}_{\vk, \sigma}^\dagger \hat{c}_{\vk, \sigma} - \sum_{\vk} \D \hat{c}_{\vk, \uparrow}^\dagger \hat{c}_{-\vk, \downarrow}^\dagger \,,
\end{align}
where $\eps$ is the electronic band dispersion and $\Delta_\vk$ the momentum-dependent superconducting order parameter. This latter is described by a complex number which satisfies the gap equation
\begin{align} \label{gapEq}
\Delta_\vk = \sum_{\vk'} V_{\vk,\vk'} \langle \hat{c}_{-\vk', \downarrow} \hat{c}_{\vk', \uparrow} \rangle \,,
\end{align}
with $V_{\vk, \vk'}$ being the (momentum-dependent) pairing interaction. It can be factorized as $V_{\vk, \vk'} = V f_\vk f_{\vk'}$, with $f_\vk$ the form factor of the superconducting order parameter: for s-wave pairing $f_\vk =1$, while for d-wave pairing $f^{d_{x^2-y^2}}_\vk = (\cos k_x - \cos k_y)/2$. Therefore, it follows from Eq.~(\ref{gapEq}) that the gap function itself can be factorized as $\D = \Delta_0 f_\vk $. \\

We write the BCS Hamiltonian using the pseudospin formalism as in Refs. \cite{PhysRevB.92.064508,PhysRevB.101.184519}, namely
\begin{align}
\hat{H}_{BCS} = \sum_\vk \mathbf{b}_\vk \cdot \hat{\mathbf{\sigma}}_\vk \,,
\end{align}
with the pseudospin vector
\begin{align}
\hat{\mathbf{\sigma}}_\vk = \frac{1}{2} \hat{\Psi}_\vk^\dagger \mathbf{\tau} \hat{\Psi}_\vk \,,
\end{align}
which is defined in Nambu-Gor'kov space, with spinor $\hat{\Psi}_\vk^\dagger = (\hat{c}_{\vk, \uparrow}^\dagger \quad \hat{c}_{- \vk, \downarrow})$ and the Pauli matrices $\mathbf{\tau} = (\tau_1, \tau_2, \tau_3)$. The pseudo-magnetic field is defined by the vector
\begin{align}
\mathbf{b}_\vk = (- \Delta' f_\vk, - \Delta'' f_\vk,  \eps) \,,
\end{align}
where $\eps = \xi_\vk - \mu$, $\xi_\vk$ being the fermionic band dispersion, $\mu$ the chemical potential. 

In the presence of an external gauge field represented by the vector potential $\mathbf{A}(t)$ coupling to the electrons, the pseudospin changes in time according to
\begin{align} \label{sigma}
\mathbf{\sigma}_\vk (t) = \mathbf{\sigma}_\vk (0) + \delta \mathbf{\sigma}_\vk (t) \,, 
\end{align}
with $\delta \mathbf{\sigma}_\vk (t) = (x_\vk (t), y_\vk (t),  z_\vk (t))$. The external electromagnetic field is included in the pseudo-magnetic field by means of the minimal substitution $\vk \rightarrow \vk - e \mathbf{A}(t)$ in the fermionic energy, resulting in
\begin{align}
\mathbf{b}_\vk (t) = (- \Delta'(t) f_\vk,  - \Delta''(t) f_\vk,  \varepsilon_{\vk - e \mathbf{A}(t)} + \varepsilon_{\vk + e \mathbf{A}(t)}) \,.
\end{align}

The Heisenberg equation of motion for the pseudospin can be written in the Bloch form, $\partial_t \mathbf{\sigma}_{k} = 2 \textbf{b}_{\textbf{k}} \times \mathbf{\sigma}_{\textbf{k}}$, providing the set of differential equations
\begin{eqnarray}
    \left\{
    \begin{array}{ll}
    \begin{split}
        \partial_t x(t) &= - (\varepsilon_{\vk - e \mathbf{A}} + \varepsilon_{\vk + e \mathbf{A}}) y(t) - \dfrac{f_\vk}{\E} \eps \delta \Delta'' (t)  \\
        &+ 2 \delta \Delta'' (t) f_\vk z(t) \,, \\
        \partial_t y(t) &= 2 \varepsilon_{\textbf{k}} x(t) + 2 ( \Delta + \delta \Delta' (t) ) f_\vk z(t) \\
        &- \ \delta \Delta' f_\vk \dfrac{\eps}{\E} + \dfrac{\Delta f_\vk}{2 \E} (\varepsilon_{\vk - e \mathbf{A}} + \varepsilon_{\vk + e \mathbf{A}} - 2 \eps) \,, \\
        \partial_t z(t) &= -2 \ \Delta f_\vk \ y(t) - \dfrac{\Delta f_\vk^2}{\E} \delta \Delta'' (t) - 2 \delta \Delta'' (t) f_\vk x(t) \,.
    \end{split}
    \end{array}
    \right. 
    \label{Blocheqs}
\end{eqnarray}
Here, for simplicity of calculations and without loss of generality, we assumed a real order parameter, therefore $\Delta'' (t= - \infty) = 0$ at  equilibrium, so that $y(- \infty) = 0$. \\
Moreover, in order to describe a quench-drive experiment, we choose the appropriate total vector potential $\mathbf{A}(t) = \mathbf{A}_{q} (t) + \mathbf{A}_{d} (t) = \mathbf{\overline{A}}_{q} (t-t_q) + \mathbf{\overline{A}}_{d} (t-t_d)$, where $ \mathbf{A}_{q} (t)$ is the quench pulse centered at time $t=t_q$, $ \mathbf{A}_{d} (t)$ is the driving field centered at $t=t_d$. The expressions we used for the modulus of the quench pulse is a gaussian-modulated wave
\begin{align} \label{Aq}
\overline{A}_q(t - t_q) = A_q \ e^{-(t-t_q)^2/\tau_q^2} \cos{(\omega_q (t-t_q))} \,,
\end{align}
while for the driving we adopted an asymmetric pulse (see Fig.\ref{plot}(b)), in order to induce an effective quench to the superconductor at the initial time:
\begin{widetext}
\begin{eqnarray} \label{Ad}
\overline{A}_d(t) =
    \left\{
    \begin{array}{l}
    \begin{split}
        & A_d \dfrac{\sin{(\omega_d (t-t_d))}}{1+(t-t_d)^2} (t-t_d) \ e^{-(t-t_d)/\tau_d^2} \quad \,, \quad \text{for} \ t \geq t_d \,, \\  
        & 0 \quad \,, \quad \text{for} \ t < t_d \,.
    \end{split}
    \end{array}
    \right.
\end{eqnarray}
\end{widetext}
Introducing the quench-drive time-delay $\Delta t = t_d - t_q$ and choosing $t_d=0$, we can rewrite $\mathbf{A}(t) = \mathbf{\overline{A}}_{q} (t + \Delta t) + \mathbf{\overline{A}}_{d} (\overline{t})$. Therefore the expressions in Eq.s (\ref{sigma})-(\ref{Blocheqs}) depend on both t and $\Delta t$. \\
The solution of Eq.~(\ref{Blocheqs}) provides the the full time-dependent pseudospin, from which the time-dependent order parameter $\Delta (t)$ can be calculated, namely
\begin{align}
\Delta_\vk (t) = V f_\vk \sum_{\vk'} f_{\vk'} \left(\sigma_{\vk, x} (t) - i \sigma_{\vk,y} (t) \right) \,,
\end{align}
where $\sigma_{\vk, x} (t)$ ($\sigma_{\vk, y} (t)$) is the $x$ ($y$) time-dependent component of the full pseudospin. The current generated by the superconductor in this quench-drive setup is given by the expression
\begin{align}
\mathbf{j}(t, \Delta t) = e \sum_{\vk} \mathbf{v}_{\vk - e \mathbf{A}(t, \Delta t)} \langle \hat{c}^\dagger_{\vk, \uparrow} \hat{c}_{\vk, \uparrow} +  \hat{c}^\dagger_{\vk, \downarrow} \hat{c}_{\vk, \downarrow}  \rangle (t, \Delta t)  \,,
\end{align}
where the electron band velocity is calculated via $\mathbf{v}_{\vk - e \mathbf{A}(t, \Delta t)} = \nabla_\vk \varepsilon_{\vk - e \mathbf{A}(t, \Delta t)}$. In particular, if we consider to measure the current generated along the $x$ direction, it can calculated by the expression
\begin{align}
j_x (t, \Delta t) = e \sum_{\vk} \dfrac{\partial \varepsilon_{\vk - e \mathbf{A}(t, \Delta t)}}{\partial k_x} \langle \hat{n}_{\vk} (t, \Delta t) \rangle  \,,
\end{align}
However, since the fundamental harmonic is dominant in this regime, while the superconducting features are visible in the nonlinear response, we are interested in the lowest order nonlinear current contribution. The first non-vanishing nonlinear term generated by the driving pulse is the third order component, which reads
\begin{align}
\mathbf{j}^{(3)} (t, \Delta t) = - 2 e^2 \sum_\vk \sum_{i=x,y,z} \mathbf{A} (t, \Delta t) \cdot \mathbf{r}_i \left( \partial_{k_i} \mathbf{v}_\vk \right) z_\vk (t, \Delta t) \,. \label{NLcurrent}
\end{align}
Here, $\mathbf{r}_i$ is the unit vector in the direction of axis $i =x,y,z$, $\mathbf{A} (t, \Delta t)$ is the total vector potential. The non-equilibrium term of the third component of the pseudospin, $z_\vk (t, \Delta t)$, contains a quadratic dependence on the full vector potential, $A^2$, and is characterized by oscillations with frequencies $2 \omega_q$, $2 \omega_d$ and $\omega_q \pm \omega_d$, as well as $2 \Delta$ due to quasiparticles' and amplitude mode excitation. The Fourier transform of the nonlinear current with respect to both the real time $t$ and the quench-drive delay time $\Delta t$, $j^{(3)} (\omega, \omega_{\Delta t})$, provides the spectrum of the generated harmonics. \\
Identifying the factor depending on the derivative of the velocity and the direction of the external field with $C_x (\vk)$, we can write the $x$ component of the third harmonic generated current as
\begin{align}
j_x^{(3)} & (\omega = 3 \omega_d, \omega_{\Delta t} = 0) = -2 \text{e}^2 A_d \sum_\vk C_x (\vk) \notag \\ 
&z_\vk (\omega = 2 \omega_d, \omega_{\Delta t} = 0) - 2 \text{e}^2 A_d \sum_\vk C_x (\vk) \notag \\
& z_\vk (\omega = 4 \omega_d, \omega_{\Delta t} = 0) \,,
\end{align}
where the two terms provide the contribution with a sum- and difference-frequency mechanism with the external driving field, respectively. We used $\omega$ as Fourier transform of the real-time variable $t$, and $\omega_{\Delta t}$ for the Fourier transform of the quench-drive delay time $\Delta t$. Analogously, the third order contribution to the fundamental harmonic is given by
\begin{align}
j_x^{(3)}(\omega = \omega_d, \omega_{\Delta t} = 0) &= -2 \text{e}^2 A_d \sum_\vk C_x (\vk) \notag \\
&  \ z_\vk (\omega = - 2  \omega_d, \omega_{\Delta t} = 0) \,.
\end{align}
Moreover, since the third component of the pseudospin has a (non-resonant) peak in frequency domain at $\omega = 2 \Delta$, it is possible to obtain other local maxima for the generated nonlinear current at $\omega = 2 \Delta \pm \omega_d$:
\begin{align}
j_x^{(3)}(2\Delta \pm \omega_d, \omega_{\Delta t} = 0) &= - 2 \text{e}^2 A_d \sum_\vk C_x (\vk) \notag \\
& z_\vk (\omega = 2 \Delta, \omega_{\Delta t} = 0 ) \,.
\end{align}
In addition to these contributions, other terms which involve the quench pulse are present, such as - among the others - the nonlinear current term
\begin{align}
j_x^{(3)}(\omega_d, \omega_{\Delta t} = 2 \omega_q ) &= - 2 \text{e}^2 A_d \sum_\vk C_x (\vk) \notag \\
& z_\vk (\omega = 0, \omega_{\Delta t} = 2 \omega_q ) \,.
\end{align}
This expression involves a sum-frequency process of two photons of the quench pulse, each with frequency $\omega_q$, embedded in the third component of the pseudospin, $z_\vk$: thus the dependence $z_\vk (\omega_{\Delta t} = 2 \omega_q)$, and it also implicitly depends quadratically on the amplitude of the quench pulse, $A_q^2$.

\subsection{Extended pseudospin model for incoherent pairs} \label{sec:extpseudo}
We now want to describe a state with a superconducting instability and characterized by the presence of pre-formed incoherent pairs, which can be identified as the origin of the pseudogap phase: in this regard, as in Ref. \cite{PhysRevLett.122.067002}, we use a new artificial equilibrium superconducting state obtained by adding a random momentum-dependent phase $\phi_\vk$ to the original Cooper pairs' state. Therefore, the strength of the pairing potential remains unchanged, as well as the number of total Cooper pairs, while the superconducting order parameter decreases due to the reduced coherence. 
According to the maximum angle $\phi_{max}$ which defines the range of the random phase $\phi_\vk$, with $\phi_\vk \in [- \phi_{max}, + \phi_{max}]$, we are able to describe different conditions of the material, from the pure superconducting phase for $\phi_{max} = 0$, to the complete loss of coherence for $\phi_{max} = \pi$.\\
We define the gap of the pure superconducting state $\Delta^{(0)}_\vk = \Delta^{(0)}_0 f_\vk$, obtained from the pure BCS gap equation, and the superconducting order parameter in the presence of incoherent pairs of the pseudogap phase as $\tilde{\Delta}^{(\phi)}_\vk = \tilde{\Delta}^{(\phi)} f_\vk$, such that  
\begin{align} \label{PGgap}
\tilde{\Delta}^{(\phi)} = V f_\vk \sum_{\vk'} f_{\vk'}^2 \dfrac{\Delta_0^{(0)}}{2 E_{\vk'}^{(0)}} e^{i \phi_{\vk'}} \,,
\end{align}
where $V$ is the same pairing strength of the original state, and with the quasiparticles' energy $E_{\vk}^{(0)} = \sqrt{\eps^2 + (\Delta^{(0)}_0)^2}$. 
The superconducting gap in the new equilibrium state can be written using the pseudospin formalism
\begin{align}
\tilde{\Delta}^{(\phi)}_\vk = V f_\vk \sum_{\vk'} f_{\vk'} \left( \tilde{\sigma}_{\vk', x} - i \tilde{\sigma}_{\vk', y} \right) \,,
\end{align}
where we have introduced the equilibrium pseudospin components
\begin{eqnarray}
    \left\{
    \begin{array}{ll}
    \begin{split}
    \tilde{\sigma}_{\vk, x} &= \sigma_{\vk, x} \cos{\phi_\vk} = f_\vk \dfrac{\Delta_0^{(0)} \cos{\phi_\vk}}{2 E_{\vk}^{(0)}} \,, \\
    \tilde{\sigma}_{\vk, y} &= \sigma_{\vk, y} \sin{\phi_\vk} = -f_\vk \dfrac{\Delta_0^{(0)} \sin{\phi_\vk}}{2 E_{\vk}^{(0)}} \,.
    \end{split}
    \end{array}
    \right.
\end{eqnarray}
Analogously to the derivation in Section II.A for the original superconducting phase, we can obtain the Heisenberg's equation of motion 
\begin{align} \label{Heis2}
\partial_t \tilde{\sigma}_\vk = \tilde{\mathbf{b}} \times \tilde{\sigma}_\vk \,,
\end{align}
with the new pseudomagnetic field defined as
\begin{align}
\tilde{\mathbf{b}} = (- \tilde{\Delta}' f_\vk, - \tilde{\Delta}'' f_\vk,  \eps) \,.
\end{align}
The solution of Eq. (\ref{Heis2}) provides the time-dependent value of the pseudospin $\tilde{\sigma}_\vk$, from which the evolution of the new order parameter $\tilde{\Delta}^{(\phi)}$ can be obtained. 
However, we notice that the complex order parameter can be written as
\begin{align}
\tilde{\Delta}^{(\phi)} = |\tilde{\Delta}^{(\phi)}| \ e^{i \theta} \,,
\end{align}
where $\theta$ is the global phase of the superconducting gap, which differs from the local phase of the Cooper pairs in momentum space, $\phi_\vk$. As a consequence, the gap equation is not self-consistent anymore (see Eq. (\ref{PGgap})) and the value of the gap is subject to some time-dependent noise due to the phase incoherence of the preformed pairs.


\subsection{Quasi-equilibrium nonlinear susceptibility}
In this section we tackle the problem of a quenched-driven clean superconductor by means of a diagrammatic approach to calculate nonlinear susceptibility: this is provided only as a tool to interpret the numerical results obtained via the pseudospin model. \\
Starting from the BCS Hamiltonian in Eq. (\ref{BCSHam}), we add the interaction of the external field that we treat perturbatively, which can be expressed as a sum of Feynman diagrams around a quasi-equilibrium condition. In particular, the current generated by the perturbed superconductor in the quasi-equilibrium condition satisfies a proportionality relation with the generalized density-density susceptibility, namely $j(\omega) \sim \chi_{\gamma \gamma} (\omega') \ A^3(\omega - \omega')$, with $\chi_{\gamma \gamma} (\omega) = \langle \tilde{\rho} \tilde{\rho} \rangle$ and $\tilde{\rho} = \sum_\vk \gamma_\vk \langle \hat{c}^\dagger_\vk \hat{c}_\vk \rangle$.\\
Since we are considering a clean superconductor, the lowest non-vanishing order of the nonlinear response is provided by the third-order nonlinear susceptibility $\chi_{\gamma \gamma}^{(3)} (\omega)$ \cite{PhysRevB.93.180507}, where $\gamma_\vk$ is the diamagnetic light-matter (vertex) interaction strength, which can be written in the effective mass approximation as $\gamma_\vk = \sum_{i,j = k_x, k_y} \partial^2_{ij} \eps$ \cite{PhysRevLett.72.396,RevModPhys.79.175}. \\
The pure quasiparticles' diamagnetic response is given by the bare density-density susceptibility, which in Nambu notation within the Matsubara formalism reads
\begin{align} \label{Raman}
\chi_{\gamma \gamma}^{(3)} (i \nu_m) &= T \sum_{\vk, i \omega_n} \gamma_\vk^2 \ \mathrm{Tr} \big[ \hat{G}(\vk, i \omega_n) \tau_3  \hat{G}(\vk, i \omega_n + \nu_m) \tau_3 \big] \,,
\end{align}
in the limit of the light momentum $\vq \rightarrow 0$. Here the Matsubara complex frequency is $i \omega_n$, and $T$ is the temperature. The Nambu-Green's function in matricial form is given by 
\begin{align} 
\hat{G} (\vk, i \omega_n) = \dfrac{1}{(i \omega_n)^2 - E_{\vk}^2} \left(
\begin{array}{cc}
    i \omega_n + \eps & - \D \\
    - \Delta^*_{\vk} & i \omega_n - \eps
\end{array}
\right) \,.
\end{align}
This function can be expressed in its spectral form as
\begin{align}
\hat{G} (\vk, i \omega_n) = - \dfrac{1}{\pi} \int_{- \infty}^{+ \infty} \text{d} \omega \ \dfrac{\hat{G}'' (\vk, \omega + i \delta)}{i \omega_n - \omega} \,,
\end{align}
where the imaginary part of the Green's function in real frequency is
\begin{align} 
\hat{G}'' (\vk, \omega) &= - \dfrac{\pi}{2 \E} \left(
\begin{array}{cc}
    \omega + \eps & \D \\
    \Delta^*_{\vk} & \omega - \eps
\end{array} \right) \cdot \nonumber \\
& \cdot \left[ \delta(\omega - \E) - \delta (\omega + \E)  \right] \,.
\end{align}
Therefore, the expression in Eq.(\ref{Raman}) can be solved obtaining
\begin{align}
\chi^{(3)}_{\gamma \gamma} (\omega) = - \sum_\vk \gamma_\vk^2 \dfrac{\Delta_\vk^2}{E_\vk^2} \dfrac{\tanh{(\beta \E / 2)}}{(2 \E + \omega + i \delta) (2 \E - \omega - i \delta)} \,. 
\end{align}
This is the pure quasiparticles' contribution, obtained neglecting the oscillations of the order parameter for small perturbations. The Higgs propagator is obtained by the random phase approximation (RPA) summation of the pairing interaction in the amplitude channel:
\begin{align}
D_{\text{Higgs}} (\vq, \omega) = - V/2 - V/2 \ \chi_{ff} (\vq, \omega) D_{\text{Higgs}} (\vq, \omega) \,,
\end{align}
which provides the expression
\begin{align} \label{Higgs}
D_{\text{Higgs}} (\vq, \omega) = \dfrac{-1}{2/V + \chi_{ff} (\vq, \omega)} \,,
\end{align}
where $\chi_{ff} (\vq, \omega)$ is given by 
\begin{align}
\chi_{ff} (\vq, \omega) &= T \sum_{\vk, i \omega_n} f_\vk^2 \ \mathrm{Tr} \big[ \tau_1 \hat{G} (\vk, i \omega_n) \tau_1 \nonumber \\
& \hat{G} (\vk, i \omega_n + i \nu_m) \big] \bigg|_{i \nu_m \rightarrow \omega + i \delta} \,.
\end{align}
We can include in the pure quasiparticles' Raman response in Eq. (\ref{Raman}) the Higgs contribution given by the propagator in Eq. (\ref{Higgs}), obtaining the full Raman response 
\begin{align}
\chi_{\gamma \Gamma} (i \nu_m) &= T \sum_{\vk, i \omega_n} \gamma_\vk \ \mathrm{Tr} \big[ \hat{G}(\vk, i \omega_n) \ \tau_3 \nonumber \\ & \hat{G}(\vk, i \omega_n + i \nu_m) \ \hat{\Gamma} (\vk, i \nu_m) \big] \,,
\end{align}
where $\hat{\Gamma} (\vk, i \nu_m)$ is the vertex' matrix which contains the corrections due to the Higgs mode. In the RPA we can identify it as
\begin{align} \label{Gammaa}
\hat{\Gamma} (\vk, i \nu_m ) &= \gamma_\vk \tau_3 - \dfrac{V}{2} \tau_1 f_\vk T \sum_{\vk', i \omega_n} f_{\vk'} \mathrm{Tr} \big[ \tau_1 \nonumber \\
& \hat{G} (\vk', i \omega_n + i \nu_m) \hat{\Gamma} (\vk', i \omega_n) \hat{G} (\vk', i \omega_n) \big] \,.
\end{align}
Additional corrections can be included in this vertex with different forms.\\
The linearized equations of motion obtained removing higher-order terms in Eq.(\ref{Blocheqs}) provide the same result of the nonlinear susceptibility calculated from the diagrammatic contributions of the density-density response including the RPA summation of the amplitude mode, responsible for the value of $\delta \Delta (\omega)$. The real part of the oscillation of the order parameter, namely the amplitude (Higgs) mode, reads in frequency space
\begin{align} \label{AM}
\delta \Delta' (\omega) = -2 \dfrac{\sum_{\vk'} \dfrac{\gamma_{\vk'} f_{\vk'}^2 \epsilon_{\vk'} \Delta_{max}}{E_{\vk'} (4 E_{\vk'}^2 - \omega^2)} }{1/V - \sum_{\vk'} \dfrac{2 \epsilon_{\vk'}^2 f_{\vk'}^2}{E_{\vk'} (4 E_{\vk'}^2 - \omega^2)}} \,.
\end{align}
The vertex in Eq.(\ref{Gammaa}) after analytic continuation of the complex Matsubara frequency can be written including explicitly the amplitude mode in Eq.(\ref{AM}) as $\hat{\Gamma} (\vk, \omega) = \gamma_\vk \tau_3 + \delta \Delta' (\omega) f_\vk \tau_1$.\\
The expression of the vertex factor $\gamma_\vk$ depends on the symmetry of the response which is measured; however, if we consider the experimental configurations with linearly polarized light along high-symmetry crystallographic directions, we can replace the factor $\gamma_\vk$ with the tensor $\gamma_{ij} (\vk) = \partial^2_{ij} \eps$, and define the corresponding susceptibility from Eq.(\ref{Raman}) as follows:
\begin{align} \label{Raman2}
\chi_{i j l m}^{(3)} (i \nu_m) &= T \sum_{\vk, i \omega_n} \gamma_{ij} (\vk) \gamma_{lm} (\vk) \ \mathrm{Tr} \big[ \hat{G}(\vk, i \omega_n) \tau_3 \nonumber \\
& \hat{G}(\vk, i \omega_n + \nu_m) \tau_3 \big] \,.
\end{align}
For example, for quench and drive pulses with nonzero components along both $x$ and $y$ directions, the third-order nonlinear response current generated along the $x$ axis will be given by
\begin{align}
j^{(3)}_x (\omega') & \propto \sum_{i,j=x,y} \chi_{ijxx}^{(3)} (\omega) A_i (\omega_1) A_j (\omega_2) \notag \\
& \delta(\omega' - \omega - \omega_1 - \omega_2) \,,
\end{align}
where $\omega'$ is obtained by energy conservation, the vector potential $A_{i,j}$ is the $i,j=x,y$-component of the total vector potential given by the sum of quench and drive. 

We notice that, in contrast to the solution of the Heisenberg's equations of motion which are valid on a general basis, the diagrammatic approach is valid for small perturbations of the superconducting order parameter $\Delta$. Therefore, when the intensity of the external pulses is such that the gap is significantly enhanced or suppressed, so that a new (transient) equilibrium value of the gap $\Delta'$ is reached, the quasi-equilibrium susceptibility calculation fails to catch all the features of the corresponding nonlinear response, requiring a full non-equilibrium calculation. \\
Moreover, this approach cannot be easily applied when pre-formed pairs or superconducting islands are present in the material, such as in the pseudogap phase of a cuprate, giving rise to long-range incoherence of the Cooper pairs. 

\begin{table}[tb]
\small
  \caption{\ This table shows the conversion of some values of the vector potential maximum intensity $A_{max}$ at a given central frequency, as used in the calculations, into the maximum value of the corresponding electric field, $E_{max}$. Here, the calculation of $E_{max}$ has been done considering the value of the lattice parameter $a = 5.4$ \AA.}
  \label{table1}
  \begin{tabular*}{0.48\textwidth}{@{\extracolsep{\fill}}lll}
    \hline
     Frequency [THz] \ & $A_{max}$ \ & $E_{max}$ [kV/cm] \\
    \hline
  4.57 & 0.4 & 22.3 \\ 
  5.09 & 1.6 & 99.3 \\ 
  7.16 & 0.2 & 17.5 \\ 
  7.16 & 0.4 & 35.0 \\ 
  7.80 & 0.8 & 76.1 \\ 
  8.28 & 0.4 & 40.4 \\ 
  11.14 & 0.8 & 109 \\ 
  12.73 & 0.8 & 124 \\ 
    \hline
  \end{tabular*}
\end{table}

\section{Numerical results} \label{results}
We now present the results obtained from the numerical implementation of the time-dependent Bloch equations and expressions for the current derived from the pseudospin model in sections \ref{sec:pseudo} and \ref{sec:extpseudo}. 
For the calculations we used the electronic band dispersion $\eps = -2 t (\cos{k_x} + \cos{k_y} + \mu /2)$, where the wave vector's components are expressed in units of the lattice constant $a$. We used the values of $t = 125$ meV for the nearest-neighbour hopping energy, chemical potential $\mu = -0.2$ in units of $t$, in order to obtain an electron occupation $n = 0.9$ as in Ref.\cite{PhysRevLett.122.067002}: the corresponding band structure is shown in Fig.\ref{plot}(a). For the d-wave order parameter with symmetry $\Delta_\vk = \Delta_{max} (\cos k_\vk - \cos k_y )/2$ we adopted the value $\Delta_{max} = 31$ meV: the calculations were performed with a summation over the full Brillouin zone with a homogeneous square sampling and a total number of k points $N_\vk = 10^6$. For the time-dependent evolution we used a time-step of $\delta t = 3 \cdot 10^{-4}$ ps, and for the quench-drive delay $\delta \Delta t = 2.5 \cdot 10^{-2}$ ps. \\
For the pulses we used a gaussian-shaped few-cycle quench and an asymmetric long-duration drive, as expressed in Eq.s (\ref{Aq}) and (\ref{Ad}) and shown in Fig.\ref{plot}(b), linearly polarized along the $(1,1)$ direction (Fig. \ref{plot}(a)), with parameters $\tau_q^2 = 0.01$ ps$^2$ and $\tau_d^2 = 5$ ps$^2$, respectively. The maximum intensity used for each pulse is provided for the corresponding vector potential in units of $\hbar / (e \ a)$, where $e$ is the electron charge and $a$ the lattice constant: the conversion to the value of the electric field for each frequency is provided in Table \ref{table1}. In Table \ref{table2} it is reported the conversion of each frequency (in THz) of the pulses used for the calculations to the energy scale (in meV). 

\begin{table}[tb]
\small
  \caption{\ This table shows the conversion of the frequency values of the pulses used for the numerical calculations from THz to the energy scale in meV.}
  \label{table2}
  \begin{tabular*}{0.48\textwidth}{@{\extracolsep{\fill}}lll}
    \hline
    Frequency (THz) & Energy (meV) \\
    \hline
 4.57 & 18.90 \\ 
 5.09 & 21.05 \\ 
 7.16 & 29.61 \\
 7.80 & 32.26 \\
 8.28 & 34.24 \\
 11.14 & 46.07 \\ 
 12.73 & 52.65 \\ 
    \hline
  \end{tabular*}
\end{table}

\subsection{Quench-drive response of the superconducting state}
We first focus on the response of a cuprate in its superconducting phase, at optimal doping: in particular, we first analyze the features of the current response as a function of both the real-time evolution and the quench-drive delay time. Then, we investigate the effect of both quench and drive pulses on the superconducting order parameter and its amplitude oscillations. 

\begin{figure}[h!]
\centering
\includegraphics[width=7cm]{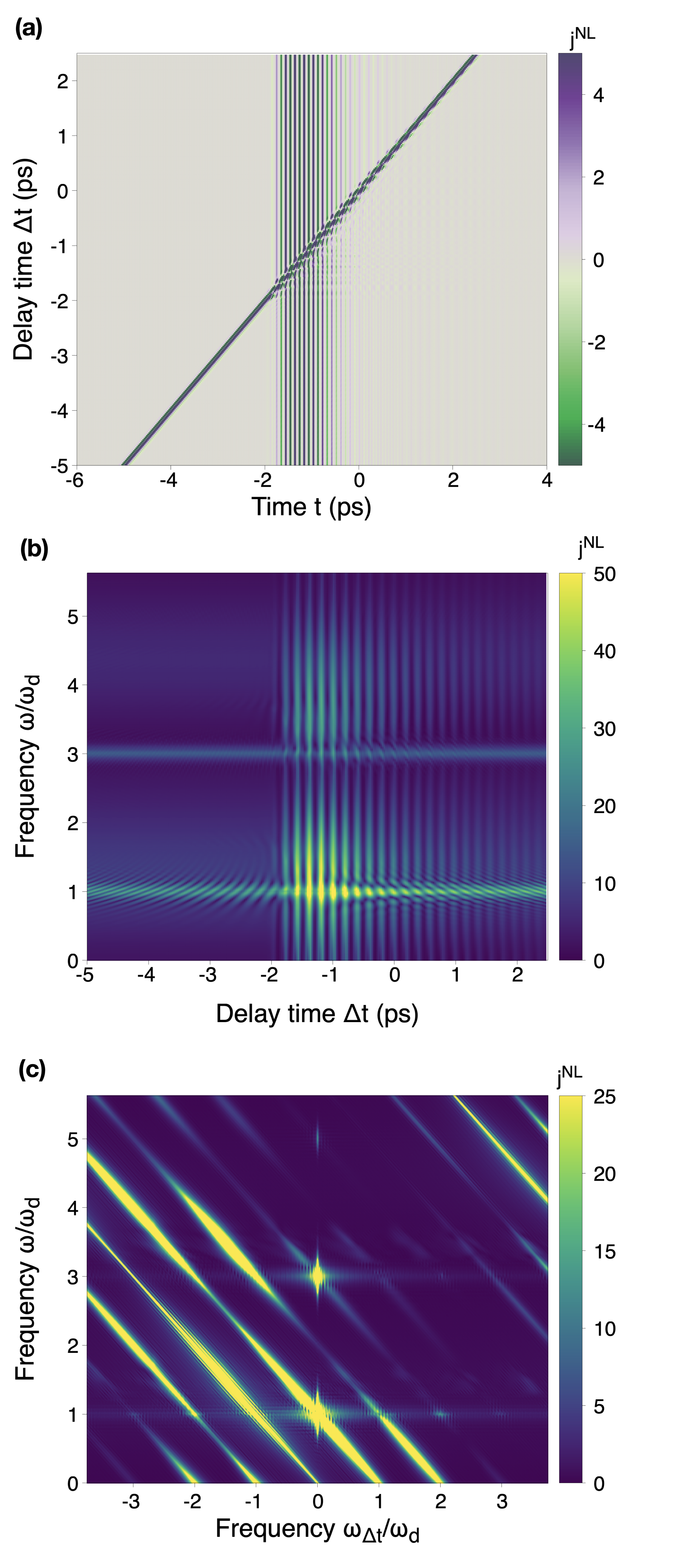}
\caption{\textbf{Two-dimensional plots of the generated nonlinear current.} \textbf{(a)} Nonlinear current response $j^{NL}$ as a function of real time $t$ and quench-drive delay time $\Delta t$. The quench frequency is $\omega_q = 7.80$ THz, the driving frequency $\omega_d = 5.09$ THz, while the maximum gap value $\Delta_{max} = 31$ meV. The maximum intensities of the pulses used was $A_q = 0.8$ and $A_d = 1.6$, respectively. The diagonal line represents the current due to the quench pulse, while the vertical lines correspond to the oscillations induced by the asymmetric driving field, which sets in at $t = -2$ ps. \textbf{(b)} The figure corresponds to the same in (a) where the horizontal axis of the real time $t$ has been Fourier transformed to $\omega$, whose values are referred to the drive frequency $\omega_d$. \textbf{(c)} 2D Fourier transform of the signal in (a), representing the nonlinear current as a function of $\omega$ and $\omega_{\Delta t}$. } \label{2Dttd1}
\end{figure}

\subsubsection{Nonlinear current generation}
We consider a quench-drive setup with both pulses linearly polarized along the diagonal direction $(1,1)$, and we analyze the nonlinear current generated along the $x$ axis, with the geometry shown in Fig. ~\ref{plot}(a). Here we used a quench with frequency $\omega_q = 7.80$ THz ($\ll 2 \Delta_{max}$) and amplitude of the vector potential $A_q = 0.8$, corresponding to an electric field $E_{max} = 76.1$ kV/cm.\\
In Fig.\ref{2Dttd1}(a) we show the nonlinear current $j^{NL}(t, \Delta t)$ generated by the superconductor as a function of the real evolution time $t$ and for different quench-drive delay times, labeled with $\Delta t$. The diagonal line represents the current induced by the quench pulse, while the vertical signal is due to the asymmetric driving field which sets in at time $t = -2$ ps. The triangular area for $0 \geq t \geq \Delta t$, for $\Delta t \in [-2, 0]$ ps is characterized by a response due to the overlap (wave mixing) of both quench and drive pulses. \\
A Fourier analysis of this plot can provide information about the harmonics present in the current itself: in Fig.\ref{2Dttd1}(b) we present the same results as a function of the frequency $\omega$, obtained by Fourier transforming the time evolution $t$, and the delay time $\Delta t$. We can clearly distinguish the fundamental and the third generated harmonic, which are modulated in the delay time as the quench is swiped with respect to the drive field. This is in accordance with previous theoretical findings on conventional s-wave superconductors \cite{puviani2021transient}. However, we also observe that faded modulated responses appears for $\Delta t > -2$ ps, both at frequencies slightly higher than $\omega_d$ and $3 \omega_d$, respectively: they also show a modulation in the delay time as the other two aforementioned harmonics. Since their appearance and intensity modulation match the time at which the driving field sets in, these features can be attributed to the wave mixing pattern due to the overlap in time of the quench and the driving field. \\
In Fig.\ref{2Dttd1}(c) the full two-dimensional Fourier transform of the nonlinear current is shown, with the frequency scales referred to the driving frequency $\omega_d$. The bright spots along the vertical axis at $\omega_{\Delta t} = 0$ correspond to the static harmonics generated by the driving field only, while the diagonal stripes are given by wave mixing of the quench and the drive pulses.

\begin{figure}[h!]
\centering
\includegraphics[width=8cm]{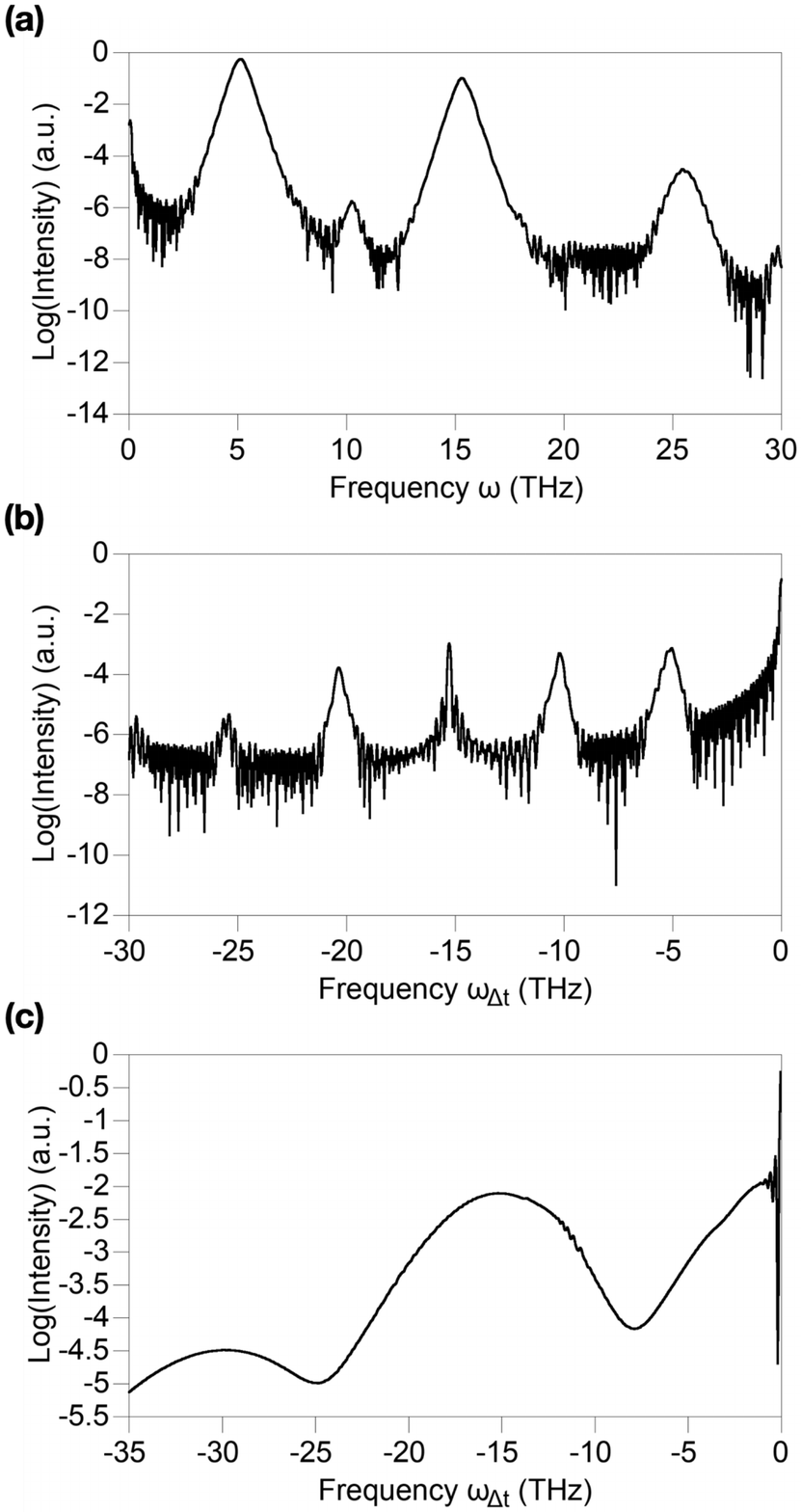}
\caption{\textbf{One dimensional plots of nonlinear current.} Plots of the nonlinear current $j^{NL}$ extracted from  Fig.\ref{2Dttd1}(c) along some relevant directions. (a) Current spectrum for $\omega_{\Delta t = 0}$ as a function of $\omega$, corresponding to a vertical cut in the two-dimensional Fourier spectrum. (b) Spectrum of the modulation of the third harmonic (TH), corresponding to a cut at $\omega = 3 \omega_d$ parallel to the $x$ axis. (c) Current intensity for $\omega = - \omega_{\Delta t} + \omega_d$, corresponding to the diagonal feature in Fig.\ref{2DNL}(a) starting from the fundamental harmonic at $\omega = \omega_q, \omega_{\Delta t} = 0$. } \label{1Dplots}
\end{figure}

\subsubsection{Transient excitation of quasiparticles and Higgs}
In order to disentangle the main features, it is useful to separately extract one-dimensional tracks along high-interest directions in this 2D spectrum: in Fig.\ref{1Dplots}(a) we highlight the spectral intensity of the nonlinear current obtained by a vertical cut at $\omega_{\Delta t} = 0$ in Fig.\ref{2Dttd1}(c). The frequency values are there provided in THz. However, despite the value of the absorption frequency $\omega_{\Delta t} = 0$, this is not completely equivalent to the case of driven superconductor in the absence of the quench pulse. In fact, a peak at $\omega \approx 10$ THz $= 2 \omega_d$ shows up: a second harmonic generation (SHG), however, cannot be generated in a driven material without breaking spatial inversion symmetry, which is not the case here. However, the presence of such SHG can be interpreted as a transient signal due to the overlap of the quench and the drive pulses: as can be seen in Fig.\ref{2Dttd1}(c), in fact, the weak peak at $\omega = 2 \omega_d, \omega_{\Delta t} = 0$ is the result of the sideband contribution of the diagonal feature given by wave mixing of the external field's components. Moreover, in addition to the fundamental and third harmonic, a fifth harmonic is present, in accordance with what previously reported in literature \cite{wang2021transient}.\\
In Fig.\ref{1Dplots}(b) we have reported the spectrum of the generated third harmonic with respect to absorption frequency $\omega_{\Delta t}$: this corresponds to track the values at $\omega = 3 \omega_d$ parallel to the horizontal axis or, equivalently, to analyze the modulation's spectrum of the third harmonic in Fig.\ref{2Dttd1}(b). In fact, the third harmonic signal is modulated in the quench-drive delay time with intensity peaks at frequencies of multiples of $\omega_d$: however, the peak at $\omega_{\Delta t} \approx 15$ THz is particularly sharp with respect to the others because of the resonance condition with the quasiparticles' and Higgs energies $2 \Delta_{max}$, being $\Delta_{max} = 31$ meV $\approx 7.5$ THz. Therefore, the dynamical modulation of the third harmonic is here enhanced by the transient excitation of quasiparticles and the Higgs mode, the latter responsible for the amplitude oscillation of the order parameter, which will be discussed in the next section. Indeed, the enhancement of the broadband quench fundamental signal by these two contributions can also be detected in Fig.\ref{1Dplots}(c), where the track of $\omega = - \omega_{\Delta t} + \omega_d$ from Fig.\ref{2Dgap}(c) is shown.

\subsubsection{Superconducting order parameter and amplitude mode}
\begin{figure}[h!]
\centering
\includegraphics[width=8cm]{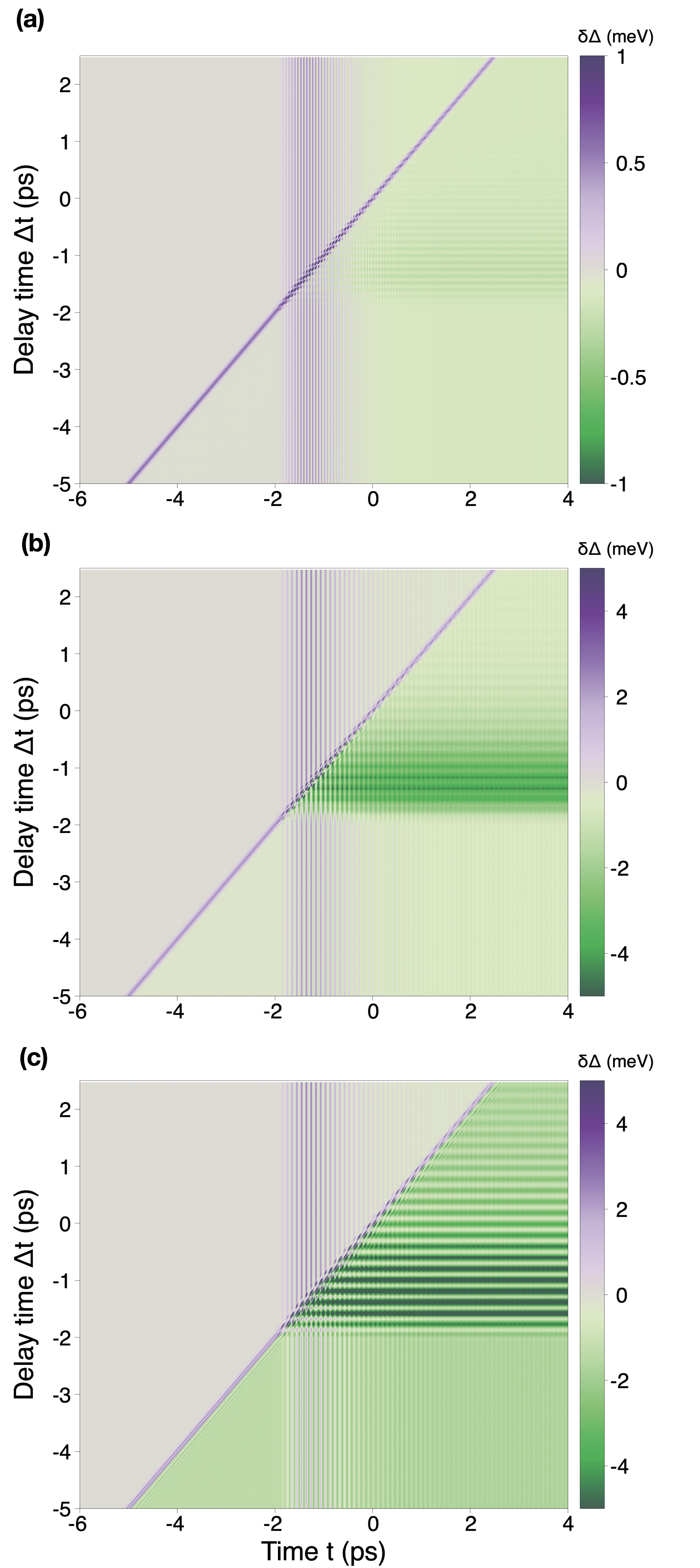}
\caption{\textbf{Two-dimensional plots of the amplitude mode.} The plots represent the amplitude mode, i.e. the variation of the order parameter $\delta \Delta (t, \Delta t) = \Delta (t, \Delta t) - \Delta_{max}$, as a function of real time $t$ and quench-drive delay time $\Delta t$, for different quench and drive pulses. (a) Quench frequency $\omega_q = 8.28$ THz, intensity $A_q = 0.4$, driving with $\omega_d = 12.73$ THz and $A_d = 0.8$. (b) Pulses' frequencies $\omega_q = 5.09$ and $\omega_d = 12.73$, intensities $A_q = 0.8$ and $A_d = 1.6$. (c) The same parameters as in (b) were used, except for the quench frequency $\omega_q = 7.80$ THz.} \label{2Dgap}
\end{figure}

Additional information can be obtained by investigating the behaviour of the superconducting order parameter by using the same two-times plots, similarly to the nonlinear current analysis. Indeed, as shown in Sections 2.1 and 2.3, the order parameter oscillates due to the pseudospins' precession in momentum space, and this collective amplitude (Higgs) mode contributes to the enhancement of the nonlinear response, shown in the previous section.\\
In Fig.\ref{2Dgap} we compare the gap variation $\delta \Delta$ for different parameters of the external fields, plotting them as a function of the time evolution $t$ and the delay time of the quench and drive pulses, $\Delta t$. \\
In Fig.\ref{2Dgap}(a) we used a quench with frequency $\omega_q = 8.28$ THz and intensity $A_q = 0.4$ and a driving pulse with $\omega_d = 12.73$ THz and $A_d = 0.8$. The oscillations of the order parameter have an amplitude $\delta \Delta_{max} < 1$ meV: they are overall positive for the duration of the quench and the drive, but they become negative for longer times until a new equilibrium value of the gap (smaller than the initial one) is reached. In Fig.\ref{2Dgap}(b) the frequency of the pulses have been set $\omega_q = 5.09$ and $\omega_d = 12.73$, while their intensities $A_q = 0.8$ and $A_d = 1.6$, respectively. Due to the higher fluence of both pulses, the oscillations of the gap are higher (up to $4-5$ meV) and the effect of the pump becomes stronger: even during the time overlap of quench drive, in fact, the gap is suppressed and characterized by an intensity modulation in the delay time $\Delta t$.
\\
In Fig.\ref{2Dgap}(c) we reduced the quench frequency to $\omega_q = 7.80$ THz, obtaining the same parameters of the calculations of the previous section on the nonlinear current. The suppression of the gap by the quench pulse is enhanced here due to the approaching of the resonance condition $\omega_q \approx \Delta_{max}$, and the intensity modulation in $\Delta t$ for $t > 2$ ps, namely when the quench and the drive overlap, is increased. \\
Therefore, we can conclude that higher fluences are associated, for higher frequencies, to a suppression of the gap and to an enhancement of its oscillations at the same time, which leads to a more intense amplitude mode and a sizeable response in the nonlinear current. On the contrary, for lower frequencies the gap is transiently enhanced during both the quench and the driving, in accordance with previous results \cite{PhysRevLett.122.067002}.

\subsection{Quench-drive response of incoherent pairs}
Similarly to the study of the pure superconducting phase of cuprates of the previous section, we now focus on the response of the same cuprate structure in the presence of reduced phase coherent giving rise to pre-formed incoherent Cooper pairs, as described in Section II.D. In particular, for our calculations we adopted a random phase within the range $[- \pi/2, \pi/2]$, which decreases the initial gap value of $\Delta_{max} = 31$ meV to $\tilde{\Delta}^{(\phi)} = 27.92$ meV. We also used the same duration and shape parameters for both the quench and the (asymmetric) drive pulse.

\subsubsection{Nonlinear current generation}
\begin{figure}[h!]
\centering
\includegraphics[width=8cm]{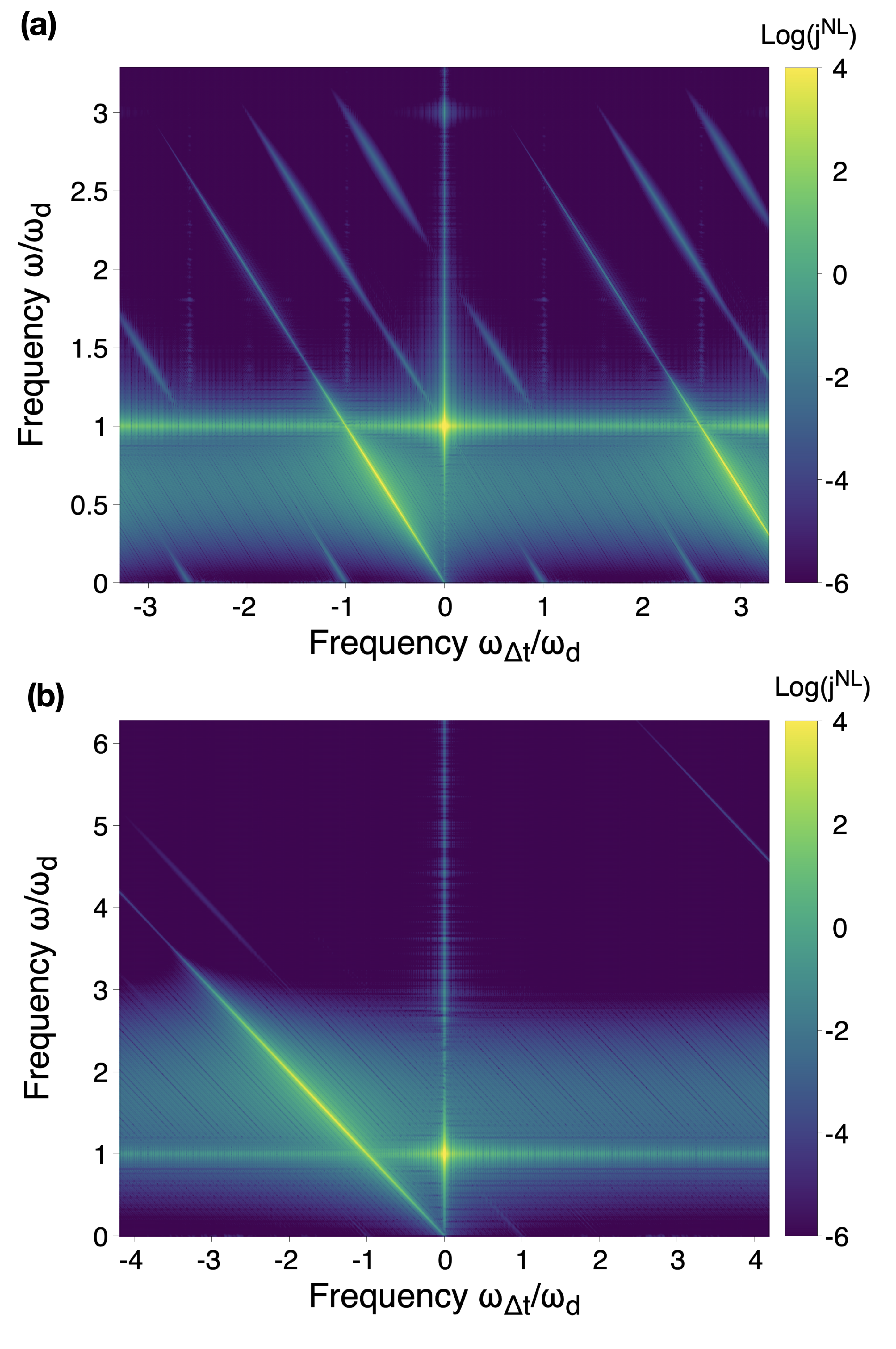}
\caption{\textbf{Two-dimensional plots in Fourier space of the generated nonlinear current.} \textbf{(a).} Nonlinear current (log scale) generated by the superconductor with quench intensity $A_q = 0.4$ and frequency $\omega_q = 7.16$ THz, driving intensity $A_d = 0.8$ and frequency $\omega_d = 11.14$ THz. \textbf{(b).} Nonlinear current (log scale) generated by the superconductor with quench intensity $A_q = 0.2$ and frequency $\omega_q = 7.16$ THz, driving intensity $A_d = 0.4$ and frequency $\omega_d = 4.57$ THz.} \label{2DNL}
\end{figure}

\begin{figure}[h!]
\centering
\includegraphics[width=8cm]{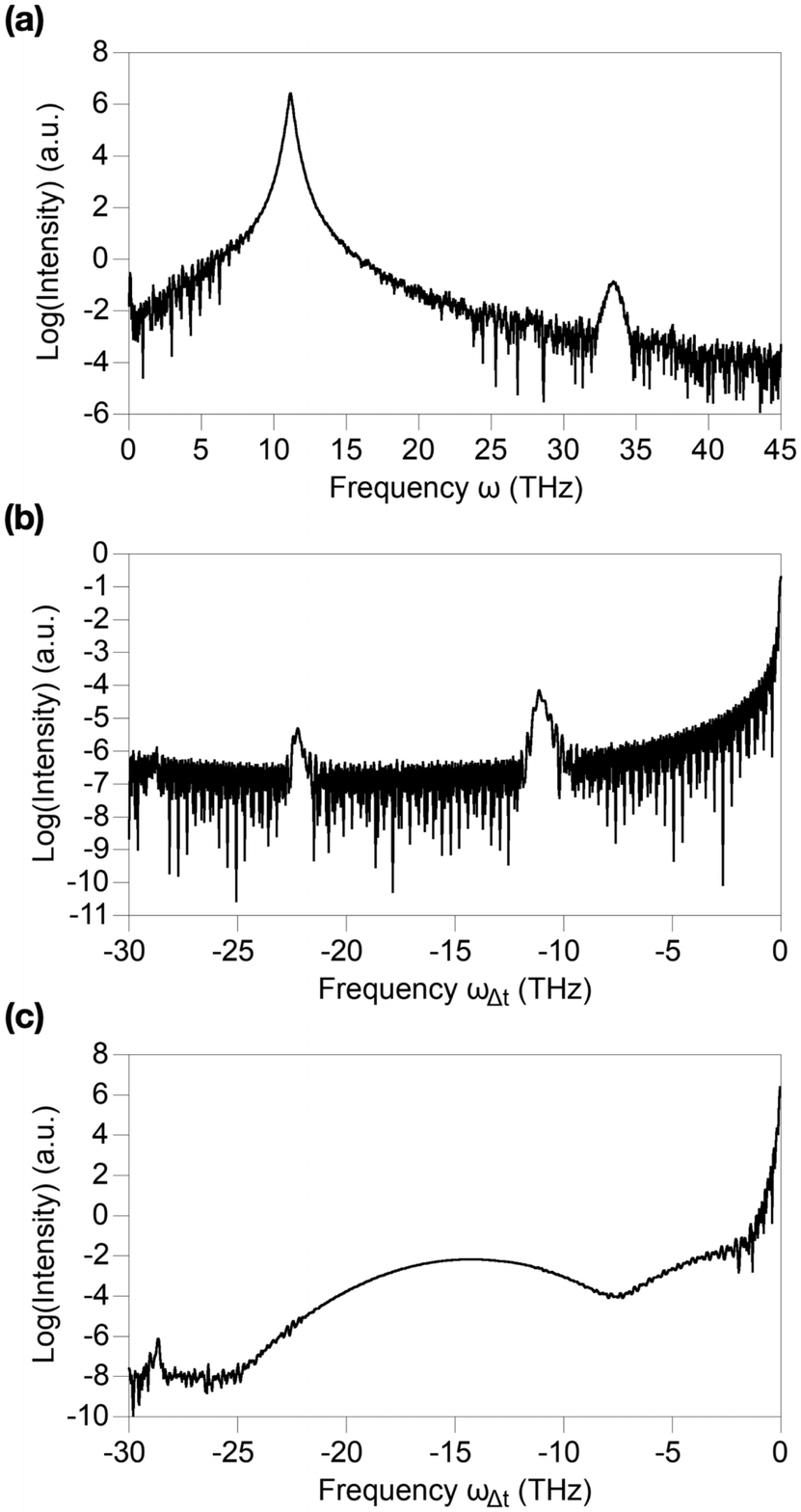}
\caption{\textbf{One dimensional plots of nonlinear current.} These plots represent the nonlinear current $j^{NL}$ in log scale extracted from  Fig.\ref{2DNL}(a) along some relevant directions. (a) Current spectrum for $\omega_{\Delta t = 0}$ as a function of $\omega$, corresponding to a vertical cut in the two-dimensional Fourier spectrum. (b) Spectrum of the modulation of the third harmonic, corresponding to a cut at $\omega = 3 \omega_d$ parallel to the $x$ axis. (c) Current intensity for $\omega = - \omega_{\Delta t} + \omega_d$, corresponding to the diagonal feature in Fig.\ref{2DNL}(a) starting from the fundamental harmonic at $\omega = \omega_q, \omega_{\Delta t} = 0$.} \label{1D2}
\end{figure}

We study here the nonlinear current generated by the superconductor with incoherent pairs under the application of a quench and a drive, with the same setup described before. In particular, we focus directly on the two-dimensional Fourier spectrum of the current, since all the other features in time domain are qualitatively similar to the pure superconducting case. \\
In Fig.\ref{2DNL}(a) we show the result for $A_q = 0.4$, $\omega_q = 7.16$ THz, $A_d = 0.8$ and $\omega_d = 11.14$ THz: while a third harmonic generated by the driving field is still present, its intensity is strongly suppressed (notice the log scale used). This is confirmed also by the one dimensional track in Fig.\ref{1D2}(a), which shows 6 orders of magnitude difference between the first and third generated harmonic. \\
In Fig.\ref{1D2}(b) the spectrum of the TH modulation is shown: while the $2 \omega_d$ and $4 \omega_d$ are still present, the peak at $2 \Delta$ is suppressed, in contrast to the result of the pure superconducting state (Fig.\ref{1Dplots}(b)). The same conclusion can be reached from Fig.\ref{1D2}(c), where it is shown the diagonal track of Fig.\ref{2DNL}(a), corresponding to $\omega = - \omega_{\Delta t} + \omega_d$, where the broadband signal of the quench pulse is visible, which is not enhanced here by the quasiparticles' and Higgs resonance in contrast to the pure superconducting case.
\\
If we further decrease the intensity of the pulses, we obtain a complete suppression of third harmonic with respect to the fundamental: this is shown in Fig.\ref{2DNL}(b), where we used $A_q = 0.2$, $\omega_q = 7.16$ THz, $A_d = 0.4$, $\omega_d = 4.57$ THz. In this case, even approaching the resonance condition $\omega_d \approx \tilde{\Delta}^{(\phi)}$, the third harmonic is not present, as well as the dynamical modulation of harmonics.\\

\subsubsection{Superconducting order parameter and amplitude mode}
\begin{figure}[h!]
\centering
\includegraphics[width=8cm]{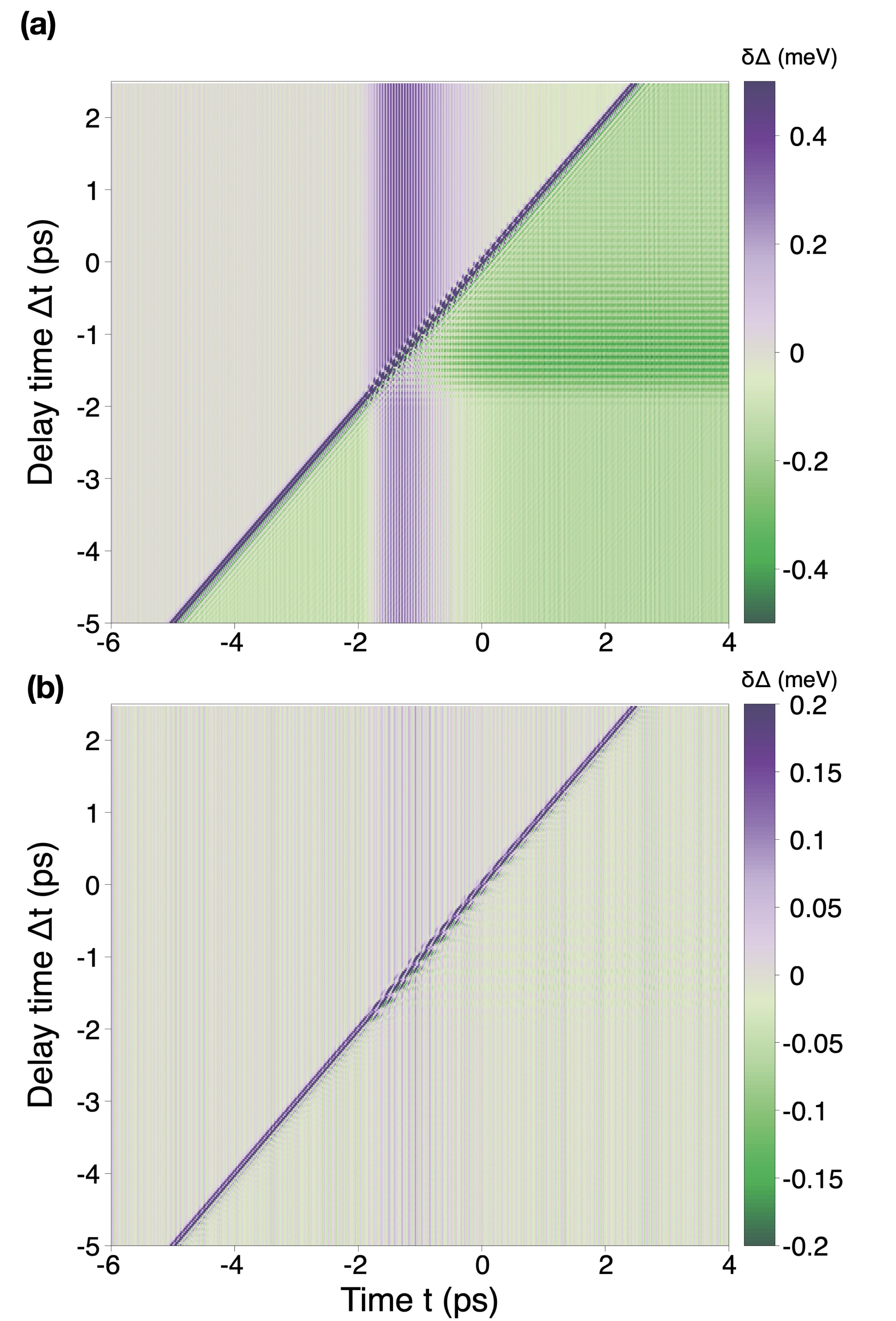}
\caption{\textbf{Comparison of the time-dependent order parameter.} Plot of the time-dependent gap change $\delta \Delta (t, \Delta t)$, referred to the equilibrium value. The results are obtained with the same parameters of the corresponding plots in Fig.\ref{2DNL}. } \label{Gap2}
\end{figure}

We now analyze the time-dependent response of the superconducting order to the quench-drive setup in the same conditions and with corresponding parameters of the nonlinear current discussed above. Indeed, the interpretation of the previous results of the nonlinear current can be supported by the study the amplitude mode of the superconducting order parameter. \\
In Fig.\ref{Gap2} we plotted the time-dependent change of the superconducting order parameter $\delta \Delta (t, \Delta t)$, corresponding to the same setup conditions of the results in Fig.\ref{2DNL}. In particular, we observe that stronger quench and drive pulses enhance the gap for the duration of the perturbation itself, except when the quench overlaps in time with the driving field, i.e. $t \in [-2, 0]$ ps (see Fig.\ref{Gap2}(a)). On the contrary, weak pulses do not enhance significantly the superconducting gap, nor activate amplitude oscillations (Fig. \ref{Gap2}(b)). As deduced from the nonlinear current response described in the previous section, the amplitude mode is suppressed by the presence of incoherence in the Cooper pairs: physically, this can be understood by the fact that instead of one coherent amplitude mode with a definite phase, we have a dispersion of modes with different phases and signs, which tend to cancel each other. Therefore, unless strong pulses increase the superconducting gap and hence the phase coherence, the Higgs and quasiparticles' mode are suppressed.

\section{Conclusions} \label{conclusion}
In this work we have studied the quench-drive spectroscopy response of cuprate superconductors, which are interesting for their anisotropic $d$-wave superconducting order parameter and their rich phase diagram. In this work we have explored two different situations: the pure superconducting state and the phase with incoherent pre-formed Cooper pairs, which can reproduce the results of the pseudogap phase, or even originate from other conditions. With this method, we have moved a step forward with respect to previous experiments and calculations, where only a pump (short quench or longer drive) of cuprates was considered. In fact, in the common pump-probe configuration, only the real time evolution of the system driven out of equilibrium by the pump was tracked by the probe and analyzed. On the contrary, we have shown that in the quench-drive setup both the short-time and the long-time pulses can act as out-of-equilibrium drive on the material, affecting both the order parameter and the nonlinear current response. Moreover, this setup has an additional temporal degree of freedom with respect to pump-probe, namely the quench-drive delay time is continuously swiped in order to gain insight of the dependence of the response on the absorption frequency $\omega_{\Delta t}$. This has allowed us to investigate the high-harmonic generation, the modulation of such harmonics in the quench-drive delay time and the role of the amplitude mode in the features of the nonlinear current. In particular, we have detected a fifth harmonic (visible at $\omega = 5 \omega_d \approx 25$ THz) generated by the superconductor in some conditions, as well as a peak at the energy $2 \Delta$, which can be attributed to quasiparticles and the amplitude (Higgs) mode, as shown by the analysis of the fluence and frequency dependence of the change of the order parameter.\\
Furthermore, we have applied quench-drive spectroscopy to a cuprate with incoherent pre-formed pairs, artificially adding a random noise phase to the Cooper pairs in momentum space. The results have shown that, even for moderate incoherence, the decrease of the superconducting gap is accompanied by the partial or total suppression of the third harmonic response, as well as the amplitude mode contribution to the nonlinear current. These results can be experimentally addressed and tested by the measurement of the transmitted electric field or the nonlinear optical conductivity.
\\
A more detailed and systematic analysis of the response of incoherent pairs and the amplitude mode of their superconducting gap will be subject of a future work. We further speculate that an extended configuration, such as a pump-pump-probe scheme, could be used to add an independent time degree of freedom on the real time evolution and the pump-probe delay time, as already used to study molecular excitations and semiconductors \cite{Cundiff2013,Giannetti2016}.

\section*{Acknowledgements}
Fruitful discussions with P. M. Bonetti, R. Haenel, S. Kaiser, M.-J. Kim, M. Monaco and D. Vilardi are thankfully acknowledged.

\end{document}